\input harvmac.tex      
\noblackbox
\input epsf.tex         

%
%
%

\def\tilde{\widetilde}
\def\bar{\overline}
\def\hat{\widehat}
\def\*{\star}
\def\[{\left[}
\def\]{\right]}
\def\({\left(}		
\def\){\right)}

%
%

\def\frac#1#2{{#1 \over #2}}
\def\inv#1{{1 \over #1}}

\def\d{\partial}

\def\2pi{\hbox{$2\pi i$}}

\def\dsl{\raise.15ex\hbox{/}\kern-.57em\partial}
\def\Dsl{\,\raise.15ex\hbox{/}\mkern-.13.5mu D}
%
%

\def\al{\alpha}

%
%
		\def\CC{{\cal C}}
		\def\CF{{\cal F}}

\def\CM{{\cal M}}

	\def\CZ{{\cal Z}}

\def\2pi{\hbox{$2\pi i$}}

\def\dsl{\raise.15ex\hbox{/}\kern-.57em\partial}
\def\Dsl{\,\raise.15ex\hbox{/}\mkern-.13.5mu D}
%
%
%
\font\numbers=cmss12
\font\upright=cmu10 scaled\magstep1
\def\stroke{\vrule height8pt width0.4pt depth-0.1pt}
\def\topfleck{\vrule height8pt width0.5pt depth-5.9pt}
\def\botfleck{\vrule height2pt width0.5pt depth0.1pt}
\def\Zmath{\vcenter{\hbox{\numbers\rlap{\rlap{Z}\kern
0.8pt\topfleck}\kern
2.2pt
                   \rlap Z\kern 6pt\botfleck\kern 1pt}}}
\def\Qmath{\vcenter{\hbox{\upright\rlap{\rlap{Q}\kern
                   3.8pt\stroke}\phantom{Q}}}}
\def\Nmath{\vcenter{\hbox{\upright\rlap{I}\kern 1.7pt N}}}
\def\Cmath{\vcenter{\hbox{\upright\rlap{\rlap{C}\kern
                   3.8pt\stroke}\phantom{C}}}}
\def\Rmath{\vcenter{\hbox{\upright\rlap{I}\kern 1.7pt R}}}
\def\Z{\ifmmode\Zmath\else$\Zmath$\fi}
\def\Q{\ifmmode\Qmath\else$\Qmath$\fi}
\def\N{\ifmmode\Nmath\else$\Nmath$\fi}
\def\C{\ifmmode\Cmath\else$\Cmath$\fi}
\def\R{\ifmmode\Rmath\else$\Rmath$\fi}

\Title{\vbox{\baselineskip12pt
\hbox{ITP/97-081}
\hbox{ISAS/90/97/EP}\hbox{hep-th/9707159}}}
{\vbox{\centerline{ Integrability of Coupled Conformal Field Theories}
\centerline{ } }}


\centerline{A. LeClair\foot{On leave from Cornell University, Newman 
Laboratory, Ithaca, NY 14853.}, 
A.W.W. Ludwig}
\medskip\centerline{Institute for Theoretical Physics}
\centerline{University of California}
\centerline{Santa Barbara, CA 93106}
\bigskip
\centerline{and}
\bigskip
\centerline{G. Mussardo}
\medskip\centerline{International School for Advanced Studies} 
\centerline{Istituto Nazionale di Fisica Nucleare} 
\centerline{and} 
\centerline{International Centre for Theoretical Physics}
\centerline{34014 Trieste}

\vskip .3in

The massive phase of two--layer integrable systems is studied 
by means of RSOS restrictions of affine Toda theories. A general
classification of all possible integrable perturbations of coupled minimal 
models is pursued by an analysis of the (extended) Dynkin diagrams. 
The models considered in most detail are 
coupled minimal models which  interpolate between magnetically coupled 
Ising models and Heisenberg spin-ladders along the $c<1$ discrete series.

\Date{7/97}
%
%
%
%
%
%

%
%
%
%
%
%
%
%
%
%
%

\def\a32{d_3^{(2)}} 
\def\qa32{ {}_q \a32 }
\def\c3{c_2^{(1)}}
\def\qc3{{}_q \c3}
\def\ck{\CC^{(k)}}
\def\sigt{{\tilde{\sigma}}}
\def\vep{\varepsilon}
\def\vept{{\tilde{\varepsilon}}}

\newsec{Introduction}

{
Since the works of 
 Belavin, Polyakov and Zamolodchikov 
on conformally  invariant  two dimensional systems and integrable
deformations thereof\ref\BPZ{A.A. Belavin, A.M. Polyakov 
and A.B. Zamolodchikov,
{\it Nucl. Phys.} {\bf B 241} (1984), 333.}\ref\rzamopert{A. B. 
 Zamolodchikov, in {\it Advanced Studies in Pure
Mathematics} {\bf 19} (1989) 641; {\it Int. J. Mod. Phys.} {\bf A3}
(1988) 743.},
and
 Andrews, Baxter and Forrester on  related integrable lattice 
solid--on--solid models\ref\ABF{G.E. Andrews,
R.J. Baxter and P.J. Forrester, {\it J. Stat. Phys.} {\bf 35} (1984),193.}, 
} much important progress has  been reported on the
classification of all possible universality classes of { two-dimensional} 
statistical models as well as on the complete control of the scaling region
nearby (see, for instance \ref\ISZ{C. Itzykson, H. Saleur and J.B. Zuber, 
{\it Conformal Invariance and Applications to Statistical Mechanics}, 
(World Scientific, Singapore 1988) and references therein.}\ref\DiFra{P.
Di Francesco, P. Mathieu and D. Senechal, {\it Conformal Field Theory}, 
Springer 1997.}\ref\GM{G. Mussardo, {\it Phys. Rep.} {\bf 218} (1992),
215}). In particular, the method  of exact relativistic scattering\ref\ZZ{A.   
B. Zamolodchikov and Al.B. Zamolodchikov, {\it Ann. Phys.} {\bf
120} (1979) 253.} and related form factor techniques\ref\KW{M. Karowsky
and P. Weisz, {\it Nucl. Phys.} {\bf B139} (1978) 445.}\ref\Smirnovbook{F. 
A. Smirnov, {\it Form Factors in Completely Integrable
Models of Quantum Field Theories} (World Scientific) 1992, and references 
therein.} has permitted  an exact solution of many models, including 
 for instance the long--standing problem of the  two--dimensional Ising model 
in a magnetic field\rzamopert\ref\DM{G.  
 Delfino and G. Mussardo, {\it Nucl. Phys.} {\bf B 455 } 
(1995), 724.}. 
{ The techniques of Exact Integrability have recently
also been  shown to be a powerful tool  for  providing
non-perturbative answers for experimentally
important strongly interacting Solid State physics 
problems\ref\FLS{P. Fendley,
A.W.W. Ludwig and H. Saleur, Phys. Rev. Lett. {\bf 74} (1995) 3005;
 Phys. Rev. {\bf B52} (1995) 8934; Statphys 19, p.137 (World Scientific,
1996)}\ref\Webb{ F.P. Milliken, C.P. Umbach and R.A. Webb, Solid State Comm. 
{\bf 97}
(1996) 309.}.}

{In this paper we exhibit 
a large class of  new integrable two-dimensional systems.
These are two planar systems (one on top of the other) 
coupled together  by  operators which lead to integrable theories (Figure 1). 

\midinsert
\epsfxsize = 5 truein
\bigskip\bigskip\bigskip\bigskip
\vbox{\vskip -.1in\hbox{\centerline{\epsffile{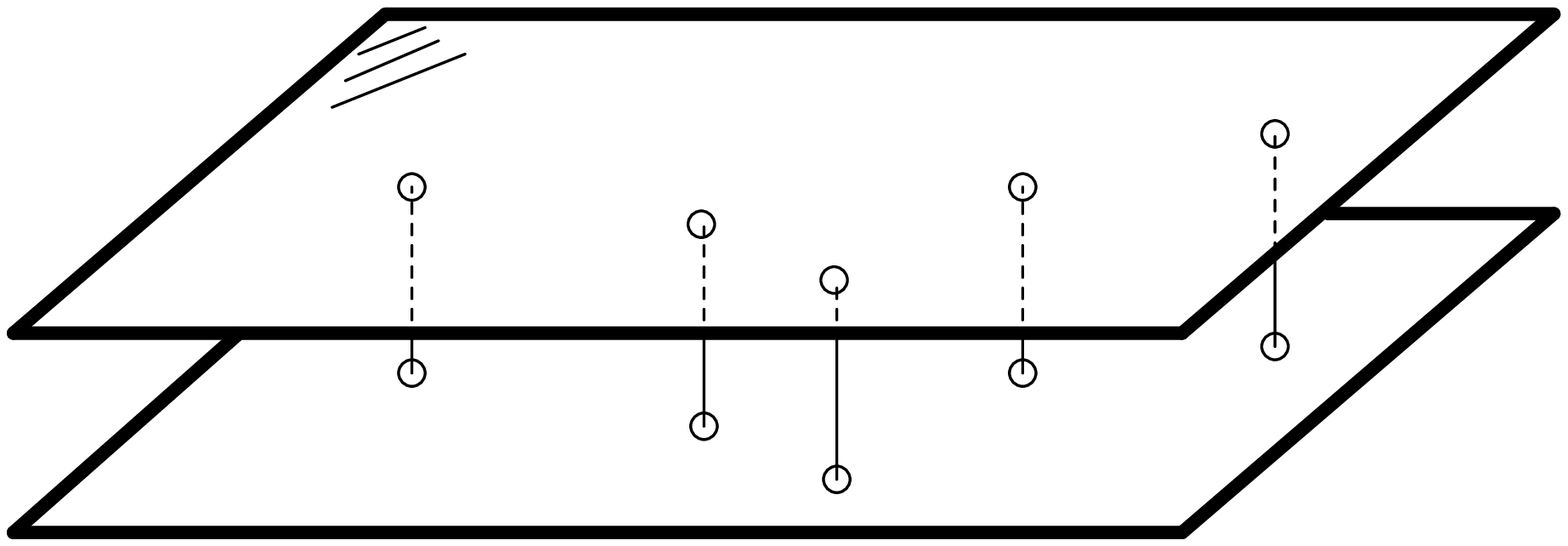}}}
\vskip .1in
{\leftskip .5in \rightskip .5in \noindent \ninerm \baselineskip=10pt
\vskip .1in
~~~~~~~~~~~~~~Figure 1. Two coupled {two-dimensional} models.
\smallskip}} 
\bigskip
\endinsert

\noindent
Our models are to be thought of as in the same category of so-called
spin-ladders
(See for instance refs. \ref\rstrong{S. P. Strong and 
A. J. Millis, Phys. Rev. Lett. 69 (1992) 2419\semi S. R. White, R. M. Noack
and D. J. Scalapino, Phys. Rev. Lett. 73 (1994) 886\semi 
 G.
Sierra,  {\it On the Application of the Non Linear Sigma Model to Spin Chains 
and Spin Ladders}, cond-mat/9610057.}\ref\rtvel{D. G. Shelton, 
A. A. Nersesyan
and A. M. Tsvelik, {\it Antiferromagnetic Spin-Ladders: Crossover
Between Spin $S=1/2$ and $S=1$ Chains}, cond-mat/9508047.}
 and references therein.) In fact, they are generalization of 
these systems.
The models we treat in most detail are 
two coupled minimal models,  interpolating  between two magnetically coupled
Ising models  and Heisenberg spin-ladders along the $c<1$ discrete series.
(The central charge of the unperturbed models ranges
from $c=1$ to $c=2$.) 
These are however only  special cases
of a much more general class of integrable models which we
identify using properties of the (extended) Dynkin
diagram of affine Lie algebras. 
These include: (i) two coupled $SO(2n)$ coset
theories, where  the central charges of the unperturbed models  range from
$c=2$ (two coupled orbifolds) to $c=2n$ (two coupled $SO(2n)_1$
current algebras), and (ii): four coupled minimal models with
unperturbed  central charges ranging from $c=2$ to $c=4$.

The integrable models studied here are bulk theories which
are  massive in the infrared.  Corresponding
integrable massless flows  in  impurity models
are studied by two of us in\ref\ALCAWWL{A. LeClair and A.W.W. Ludwig,
`` Minimal Models with Integrable Local Defects'', ITP-97-080, preprint.}.
These are generalizations of models  which have recently
attracted much attention in Condensed Matter physics, such as in
the context of  point contacts in the fractional
quantum Hall effect, and Impurities in  Quantum Wires (see e.g.
\ref\KF{ C.L. Kane and M.P.A. Fisher, Phys. Rev. {\bf B46} (1992) 15233}\FLS).

In this paper the emphasis is on the aspects coming from the integrability 
of the inter-layer coupling and on the exact results which follow. We will 
show, in particular, that the on--shell dynamics of such systems admits 
a  description in terms of an exact scattering theory. The exact
scattering amplitudes as well as the exact spectrum of excitations can be
computed by employing the RSOS reduction scheme based on the quantum 
symmetries of the models\ref\rrsm{N.  
Yu. Reshetikhin and F. Smirnov, 
{\it Commun. Math. Phys.} {\bf 131} (1990), 157.}\ref\rbl{D. Bernard and
A. LeClair, {\it Nucl. Phys.} {\bf B 340} (1990) 721;  C. Ahn, D. Bernard
and A. LeClair, {\it Nucl. Phys.} {\bf B 346} (1990) 409; A. LeClair, 
Phys. Lett. 230B (1989) 103.}. 
An important representative of the class of the models analyzed in this 
paper consists of the two--layer Ising system coupled together by their
magnetization operators $\sigma_1$ and $\sigma_2$.
 A simple mean--field analysis 
indicates that in this case the interaction between the two 
layers drives the system into a massive phase. We will determine the exact
dynamics of this model by providing the spectrum of the massive 
excitations of this model as well as all their $S$--matrix amplitudes. 
}

The paper is organized as follows: in Section 2 we analyze a particular 
integrable coupling between two minimal models of conformal
field theory (CFT), the latter being regarded 
as a coset construction on $SU(2)$. In Section 3 we study 
the integrability of coupled conformal field theories under a more 
general setting based on (Affine) Toda Field Theory. Finally, in Section 4 
the spectrum and the $S$-matrix of (magnetically) coupled minimal models 
are worked out explicitly. In Section 5 the conclusions of this 
work are presented.

\newsec{Coupled Minimal Models and ${}_q d_3^{(2)}$ and ${}_q c_2^{(1)}$ 
Affine Lie Algebras} 

Let $\ck$ denote the minimal unitary conformal field theory (CFT) with 
central charge 
\eqn\eIo{ c_k= 1-\frac{6}{(k+2)(k+3)} ~~~,} 
 $k=1,2,...$.  
These models have local primary fields
$\sigma = \Phi_{1,2} $,   $\sigt = \Phi_{2,1}$, $\vep = \Phi_{1,3}$,
and $\vept = \Phi_{3,1}$  with 
scaling dimension:
\eqn\eIi{\eqalign{
{\rm dim} \( \sigma  \)  &= 2 \Delta_{\sigma} =  
\inv{2} \( \frac{k}{k+3} \)  
~~~~~~~~
{\rm dim} \( \sigt \)  = 2 \Delta_{\sigt} = \inv{2} \( \frac{k+5}{k+2} \) 
\cr
{\rm dim} \( \vep \) &= 2 \Delta_{\vep} = 2 \( \frac{k+1}{k+3} \) 
~~~~~~~~
{\rm dim} \( \vept \) = 2 \Delta_{\vept} = 2 \( \frac{k+4}{k+2} \)  .\cr}}
(Here, $\rm dim$ refers to the sum of left and right conformal dimension.)
We define four infinite series of models $\CM_k^{\sigma}, \CM_k^{\sigt}$, 
$\CM_k^{\vep}$ and $\CM_k^{\vept}$ by coupling two copies of $\ck$ via the
operators $\sigma, \sigt$, $\vep$, $\vept$. This is described by an action 
which perturbs the tensor product of the two CFT's: 
\eqn\eIii{
{\cal A} = {\cal A}_{\ck_1} + {\cal A}_{\ck_2} + \lambda \int d^2 x 
~ \Phi_1 \Phi_2 , }
where the subscripts refer to copy 1 or 2 of $\ck$, and
$\Phi = \sigma, \sigt, \vep$ or $\vept$.  

The models $\CM_k^{\sigma}$ and $\CM_k^{\sigt}$ are characterized
by relevant perturbations for all $k$. The models $\CM_k^\vep, 
\CM_k^\vept$ on the other hand are irrelevant perturbations, except for
$\CM_1^\vep$ ({}from the dimensions \eIi, one sees in this case that
$\CM_1^{\sigt} = \CM_1^{\vep}$). The latter is a strictly marginal 
perturbation corresponding to the Ashkin-Teller model: we have then a 
{ line
of fixed points described by 
} the coupling constant $\lambda$. 
With the appropriate choice of sign of $\lambda$, the models
$\CM_k^{\sigma,\sigt}$ are massive field theories. The other models
perhaps describe the infrared limit of an integrable flow from a model 
with higher central charge in the ultraviolet. We are only concerned in
this paper with the massive models, however we will continue to point out
where the models $\CM_k^{\vep, \vept}$ reside in the algebraic 
classification. Also, this information may be useful for coupled 
non-unitary minimal models.  

One approach to integrable perturbations of minimal models and other
coset conformal field theories is based on quantum group restrictions
of affine Toda theories\ref\rhollow{T. Hollowood and
P. Mansfield, {\it Phys. Lett.} {\bf B 226} (1989), 73.}\ref\reg{T. 
Eguchi and S.K. Yang, {\it Phys. Lett.} {\bf B 224}
(1989), 373.}\rrsm\rbl. Remarkably, the same 
approach can be applied here to classify the possible integrable
perturbations of coupled minimal models. Let us see how this can be 
achieved. 

It is well known that the minimal models $\ck$ of conformal field theory 
have a description in terms of a scalar field with background charge 
\ref\DF{Vl.S. Dotsenko and V.A. Fateev, {\it Nucl. Phys.} {\bf B 240} [FS12]
(1984), 312.}. Thus two copies of $\ck$ can be represented with two scalar
fields $\phi_1 ,\phi_2$, each with the appropriate background charge to
give the requisite central charge and the conformal spectrum. In the
affine-Toda theory approach to perturbed conformal field theory, one starts
with a Toda theory on a finite Lie group $g$, then identifies the
perturbation with an affine extension of $g$ to $\hat{g}$. For our 
problem, the conformal field theory $\ck\otimes \ck$ is represented
with two scalar fields, thus the rank of $\hat{g}$ must be three. The 
other requirement of $\hat{g}$ is that when the root associated with the
perturbation is omitted, the resulting non-affine Toda theory must
be an $su(2) \otimes su(2)$ Toda theory in order to represent 
$\ck \otimes \ck$. Otherwise stated, if two coupled minimal models
can be described by a quantum group restriction of the affine Toda
theory $\hat{g}$, then the Dynkin diagram of $\hat{g}$ must contain
3 nodes, and the removal of one node must leave two decoupled nodes
with roots of the same length.  Referring to the known classification
of affine Lie algebras \ref\rgod{P. Goddard and D. Olive, {\it Int. J. Mod.
Phys.} {\bf A 1} (1986), 303.}, the only possibilities are $\c3$ and 
$\a32$. Neither of these are simply laced. The Dynkin diagrams for 
these algebras are shown in Figure 2. Removing the middle node leaves
$su(2) \otimes su(2)$, thus it is the middle node that will be  associated with
the perturbation. This is in contrast to the usual application of affine
Toda theory to perturbed coset theories, where there the extended affine root
$\alpha_0$ is associated with the perturbation.   

\midinsert
\epsfxsize = 3 truein
\bigskip\bigskip\bigskip\bigskip
\vbox{\vskip -.1in\hbox{\centerline{\epsffile{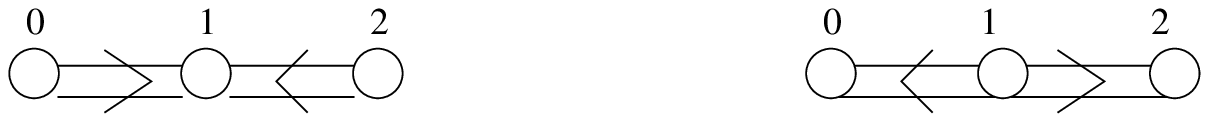}}}
\vskip .1in
{\leftskip .5in \rightskip .5in \noindent \ninerm \baselineskip=10pt
Figure 2.
Dynkin diagrams for the algebras $c_2^{(1)}$ and $d_3^{(2)}$ respectively.
\smallskip}} \bigskip
\endinsert

\def\alv{{\vec{\alpha}}}
\def\phiv{{ \vec{\phi} }} 

The affine Toda theories { associated with the Dynkin diagrams of Fig. 2}
 are defined by the action 
\eqn\eIi{
{\cal A} = \inv{4\pi} \int d^2 x \( \inv{2} \d_\mu \vec{\phi} \cdot 
\d^\mu \phiv + \lambda \sum_{\alpha_j } 
e^{-i \beta \alv_j \cdot \phiv} \) ~~~,  } 
where here
$\phiv = (\phi_1 , \phi_2 )$, $\alpha_j \in \{ \alpha_0, \alpha_1 , 
\alpha_2 \}$ are simple roots of the affine algebra, and $\beta$ is
a coupling. For a general  affine Lie algebra $\hat g$,
 ${\vec \phi}$ has ${\rm rank}(\hat{g}) -1 $ 
components and the sum runs over all
simple roots of $\hat g$.

For $\c3$, one can chose $\alv_0^2 = \alv_2^2 = 2$, $\alv_1^2 = 1$. 
The Cartan matrix $K_{ij} = 2 \alv_i \cdot \alv_j / \alpha_j^2 $ is
\eqn\Iii{
K = \left( 
\matrix{~2&-2&~0\cr -1&~ 2 & -1 \cr ~0&-2&~2 \cr } \right)~ . } 
We will also need $\alv_0 \cdot \alv_1 = \alv_1 \cdot \alv_2 = -1$. 

The algebra $\a32$ is the dual of $\c3$ under the transformation 
$\alv \to 2 \alv/\alpha^2 $. This duality is the usual one that
exchanges the orientation of the arrows of the Dynkin diagram and 
takes $K$ into its transpose. For $\a32$, one then has 
$\alv_0^2 = \alv_2^2 = 2$, $\alv_1^2 = 4$ and 
$\alv_0 \cdot \alv_1 = \alv_2 \cdot \alv_1 = -2$.     

We identify the $\al_0$ and $\al_2$ terms in the Toda potential with
the conformal field theory $\ck \otimes \ck$, which requires these
operators to have left and right conformal dimension equal to $1$. 
This can be accomplished by turning on a background charge $\vec{\gamma}$ 
with modified energy momentum tensor
\eqn\eIiii{
T = -\inv{2} \d_z \phiv \cdot \d_z \phiv + i \sqrt{2} \vec{\gamma} 
\cdot \d_z^2 \phiv ~~~. } 
We take $\vec{\gamma} = \gamma ( \alv_0 + \alv_2 )$, which leads to
the central charge $c = 2(1-48 \gamma^2 )$.  Identifying $c = 2 c_k$, 
one fixes the parameter $\gamma$ to be 
\eqn\eIiv{
\gamma = \inv{  \sqrt{8 (k+2) (k+3) } } ~~~.}  

The chiral dimension of the exponential operators are then given by  
\eqn\eIv{
\Delta \( e^{-i\beta \alv \cdot \phiv } \) = 
\beta^2 \alv^2 /2 + \sqrt{2} \beta \alv \cdot \vec{\gamma} ~~~. } 
Imposing that $e^{-i\beta \alv_0 \cdot \phiv} $ and 
$e^{-i\beta \alv_2 \cdot \phiv} $ have dimension $1$ leads to the
equation 
\eqn\eIvi{
1 = \beta^2 + 2 \sqrt{2} \beta \gamma ~~~, } 
with two solutions:
\eqn\eIvii{
\beta_+ = \sqrt{ \frac{k+2}{k+3} } ~~~, ~~~~~~
\beta_- = - \sqrt{ \frac{k+3}{k+2} } ~~~. } 
Finally, once $\beta$ and $\gamma$ are fixed we can identify the
chiral dimension of the perturbation as
$\Delta_{\rm pert} = \Delta(e^{-i\beta \alv_1 \cdot \phiv} ) $. 
For $\c3$ one finds $\Delta_{\rm pert} = 2 \Delta_\sigma$ for
$\beta_+$ and $\Delta_{\rm pert} = 2 \Delta_\sigt$ for $\beta_-$.   
For $\a32$ one finds $\Delta_{\rm pert} = 2 \Delta_\vep$ for
$\beta_+ $ and $\Delta_{\rm pert} = 2 \Delta_\vept$ for
$\beta_-$. We summarize these results by listing below the model
and its associated affine Toda theory and coupling:
\eqn\eIviii{
\eqalign{
\CM_k^\sigma ~ &: ~~~~~ \c3 ~ {\rm affine ~ Toda ~  with } ~ 
\beta = \beta_+ ~~~; \cr
\CM_k^\sigt ~ &: ~~~~~ \c3 ~ {\rm affine ~ Toda ~ with } ~ 
\beta = \beta_- ~~~; \cr
\CM_k^\vep ~ &: ~~~~~ \a32 ~ {\rm affine ~ Toda ~ with } ~ 
\beta = \beta_+ ~~~; \cr
\CM_k^\sigt ~ &: ~~~~~ \a32 ~ {\rm affine ~ Toda ~ with } ~ 
\beta = \beta_- ~~~.\cr
}}
With this identification, the spectrum and S-matrices of the models 
can be obtained as quantum group restrictions of the affine Toda theory. 
One must bear in mind that the affine Toda theory based on
$\hat{g}$ possesses the dual quantum affine symmetry ${}_q \hat{g}^\vee
$\ref\rnlc{D. Bernard and A. LeClair, {\it Commun. Math. Phys.} 
{\bf 142} (1991), 99.}\ref\rfeld{G. Felder and A. LeClair, 
{\it Int. J. Mod. Phys.} {\bf A 7} (1992), 239.} with 
\eqn\eeq{
q = e^{-i\pi/\beta^2} . }
Thus the (restricted) quantum symmetries of the models are as follows:
\eqn\eIix{
\eqalign{
\CM_k^\sigma ~ &: ~~~~~ \qa32 ~ {\rm symmetry}, ~~~ q 
= - e^{-i\pi/(k+2)} ~~~; \cr
\CM_k^\sigt ~ &: ~~~~~ \qa32 ~ {\rm symmetry}, ~~~  q 
= - e^{+i\pi/(k+3)} ~~~; \cr
\CM_k^\vep ~ &: ~~~~~ \qc3 ~ {\rm symmetry}, ~~~  q 
= - e^{-i\pi/(k+2)} ~~~; \cr
\CM_k^\vept ~ &: ~~~~~ \qc3 ~ {\rm symmetry}, ~~~  q 
= - e^{+i\pi/(k+3)} ~~~.\cr
}}

Vaysburd first established the integrability of the models 
$\CM_k^{\sigma,\sigt}$ directly as perturbations of cosets 
\ref\rvays{I. Vaysburd, {\it Nucl. Phys.} {\bf B 446} (1995), 387.}, 
by using the counting arguments of Zamolodchikov \rzamopert. The CFT 
$\ck$ can be formulated as the coset $\ck = SU(2)_k \otimes SU(2)_1 
/SU(2)_{k+1}$, where $SU(2)_k $ is the $SU(2)$ current algebra at 
level $k$\ref\rgko{P. Goddard, A. Kent and D. Olive,
Phys. Lett. {\bf 152B} (1985) 105.}. 
Using the fact that $SU(2)_k \otimes SU(2)_k = SO(4)_k $, one 
has
\eqn\eIiii{
\ck \otimes \ck = \frac{ SO(4)_k \otimes SO(4)_1 }{SO(4)_{k+1}} ~~~.}
Thus, the models $\CM_k^{\sigma, \sigt}$ can be formulated as 
perturbations of the $SO(4)$ cosets by operators of dimension 
$2 \cdot {\rm dim} \( \sigma, \sigt  \)$. These coset 
perturbations are not the generic ones which are integrable
for arbitrary Lie algebras where the perturbing field is
associated with the adjoint representation \rbl, and in the affine
Toda approach are associated with the affine root $\alv_0$; 
rather the perturbing fields here are associated with the vector
representation. As explained in \rvays, the latter corresponds to a 
different way of affinizing $SO(4)$ to yield the affine algebras 
$\a32, c_2^{(1)}$.

\newsec{General Scheme and Other Examples}
\def\hd{h^*}
\def\alv{\vec{\alpha}} 
\def\phiv{\vec{\phi}} 

\subsec{Affine Toda Theories for Coupled Conformal Field Theories}

The construction of the last section is just an example of a more
general one for studying integrability of coupled conformal field
theories based on affine Toda theories.  Let $\hat{g}$ 
denote an affine Lie algebra and $\{ \alpha (\hat{g} ) \}$ its simple
roots, $\{ \alv_0, \alv_1 , ..., \alv_r \} $.  In the Dynkin
diagram of $\hat{g}$, we identify one node and its associated root
as the perturbation and denote this root as $\alv_{pert} \in 
\{ \alpha (\hat{g} ) \} $.  We further require that upon 
removing the node $\alv_{pert}$ we are left with two decoupled
Dynkin diagrams representing $g_1 \oplus g_2$, where $g_1$ and
$g_2$ are finite dimensional, simply laced Lie algebras.  By
chosing the background charges appropriately, the conformal field
theory corresponds to two decoupled conformal Toda theories based on
$g_1$ and $g_2$, and because these are simply laced, these can
represent the coset theories of the $g_1$ and $g_2$ current algebras
\ref\rBG{A. Bilal and J.-L. Gervais, Phys. Lett. B206 (1988)
412.}\ref\rFL{V. Fateev and S. Lukyanov, Int. J. Mod. Phys. A3 (1988)
507.}. The perturbation term $\exp (-i \beta \alv_{pert} \cdot \phiv)$ 
is the one which couples the two conformal field theories, and its
dimension is fixed once the background charge is fixed.   

Normally, one choses $\alv_{pert} = \alv_0$, which is the negative
of the highest root, and always occurs at an end of the Dynkin diagram. 
This well-known case describes the perturbation of a single coset theory
since here $g_1 = g$ and $g_2$ is empty. In this case, the background 
charges require that one begin with the S-matrices  of the unrestricted
Toda theory in the homogeneous gradation, since it is in this gradation
that the ${}_q g $ invariance is manifest (See e.g. \rfeld .). For the new
cases we are considering, the background charges are different, and one
must first transform the S-matrices to the appropriate gradation where the
${}_q g_1 \oplus {}_q g_2 $ symmetry is manifested, before doing the 
restriction. As far as the spectrum and S-matrices are concerned, this is
the main dynamical difference between models with $\alv_{pert} = \alv_0$ and 
$\alv_{pert} \neq \alv_0$.    
{ We now discuss two examples of this construction.}

\subsec{Coupled $SO(2n)$ Cosets and ${}_q d_{2n}^{(1)}$ Affine Algebras} 

\def\dn{d_{2n}^{(1)}} 
\def\gv{\vec{\gamma}} 
\def\rv{\vec{\rho}} 

Let us begin with the Toda theory based on the affine algebra
$\dn$, which is the standard affinization of $d_{2n} = so(4n)$. 
Its Dynkin diagram is shown in Figure 3. If one removes the central 
node on the string, the diagram decouples into two $d_n = so(2n)$ Lie
algebras. Thus, if we identify $\alv_{pert} = \alv_n$, the $\dn$ 
affine Toda theory can be used to describe two coupled 
$so(2n)$ cosets. We denote by $\{ \alv^{(1,2)} (d_n) \}$ the 
simple roots for copies $1$ and $2$ of $so(2n)$ so that
\eqn\eroots{
\{ \alv (d_{2n}^{(1)} ) \} =  
\{ \alv^{(1)} (d_n) \} + \{ \alv^{(2)} (d_n) \} + \alv_n ~~~. } 
\midinsert
\epsfxsize = 3in
\bigskip\bigskip\bigskip\bigskip
\vbox{\vskip -.1in\hbox{\centerline{\epsffile{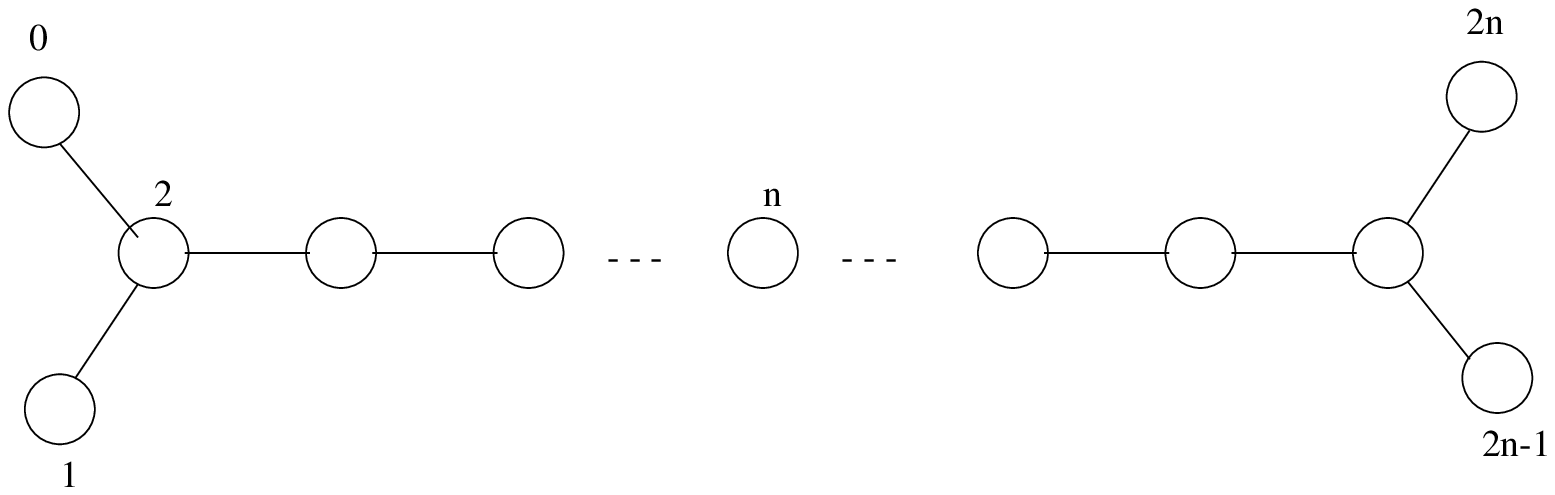}}}
\vskip .1in
{\leftskip .5in \rightskip .5in \noindent \ninerm \baselineskip=10pt
~~~Figure 2. Dynkin diagram for the affine algebra $d_{2n}^{(1)}$. 
\smallskip}} \bigskip
\endinsert
Let $\CC^{(k)}$ denote the coset CFT
\eqn\efouri{
\CC^{(k)}_n  = \frac{SO(2n)_k \otimes SO(2n)_1 }{SO(2n)_{k+1} } ~~~,}
with central charge
\eqn\efourii{
c_k^n = n \( 1 - 
\frac{\hd (\hd + 1)}{(k+ \hd)(k+\hd + 1)}\)~~~,}
where the dual Coxeter number of $so(2n)$ is $\hd = 2n-2$. The $\dn$ 
affine Toda theory contains $2n$ scalar fields  { combined
into } the vector $\phiv$.  
We let the energy momentum tensor take the form \eIiii, with central 
charge $c = 2n - 24 \vec{\gamma}^2$. The background charge $\gv$ is 
chosen such as to represent two decoupled $\CC^{(k)}_n$ theories, 
\eqn\efouriii{
\gv = 2 \gamma \( \rv_1 + \rv_2 \) , }
where $\rv_{1,2}$ are the Weyl vectors for copies $1$ and $2$ of 
$so(2n)$, namely, $\rv_{1,2} = \sum_{i=1}^n \vec{\mu}_i^{(1,2)}$, 
where $\vec{\mu}_i^{(1,2)} \cdot \alv_j^{(1,2)} = \delta_{ij}$.
This implies  
\eqn\efouriv{
\rv_{1,2} = \sum_{i=1}^n K_{ij}^{-1} \alv_j^{(1,2)}~~~, }
where $K$ is the Cartan matrix of $so(2n)$. The Weyl vectors satisfy
\eqn\efourv{
\rv_1 \cdot \rv_2 = 0, ~~~~~~(\rv_1)^2 = (\rv_2)^2 = n \hd (\hd+1)/12 ~~~.}
Identifying $c$ with $2 c_k^n$, one requires 
\eqn\efourvi{
\gamma = \inv{ \sqrt{8 (k+\hd) (k+\hd + 1 )} }~~~. }
Next we require that the terms in the Toda potential $\exp (-i\beta \alv
\cdot \phiv )$ with $\alv$ a simple root of either copy of 
$so(2n)$ to have conformal dimension equal to 1.  
This gives the equation \eIvi, with solutions 
\eqn\efourvii{
\beta_+ = \sqrt{ \frac{k+\hd}{k+\hd + 1} }~~~, ~~~~~
\beta_- = - \sqrt{ \frac{k+\hd +1 }{k+\hd } }
~~~.}
The dimension of the perturbation follows from \eIv\ and 
$(\rv_1 + \rv_2 ) \cdot \alv_{pert} = - \hd$:
\eqn\efourviii{\eqalign{
\Delta_{pert} &= \Delta^+_{pert} = \frac{k}{k+ \hd +1} ~~~, 
~~~~~~~~ {\rm for} ~~ \beta = \beta_+ ~~~;\cr
\Delta_{pert} &= \Delta^-_{pert} = \frac{k + 2\hd + 1}{k+ \hd} ~~~, 
~~~~~~~~ {\rm for} ~~ \beta = \beta_-  ~~~. \cr}}

Let us now interpret these models. The CFT $\CC^{(k)}_n$ has two 
primary fields $\Phi^e$ and $\Phi^h$, which are associated with vector
representations of $so(2n)$, with chiral scaling dimension 
\eqn\efourix{
\Delta (\Phi^h ) = \inv{2} ~\Delta_{pert}^+ ~~~, ~~~~~~~~~~
\Delta (\Phi^e ) = \inv{2} ~\Delta_{pert}^-  ~~~. }
It was shown by Vaysburd that the following perturbations of 
a single copy of $\CC^{(k)}_n$ are integrable:
\eqn\efourx{
{\cal A}^{e,h} = {\cal A}_{\CC^{(k)}_n} + \lambda \int d^2 x ~ \Phi^{e,h} . }
{}From \efourix\ one sees that our models correspond to two 
$\CC^{(k)}_n$ theories coupled by these operators:
\eqn\efourxi{
{\cal A} = {\cal A}_{\CC^{(k)}_n \otimes \CC^{(k)}_n } + \lambda 
\int d^2 x ~\Phi_1^{e,h} ~\Phi_2^{e,h} ~~~.} 

To summarize, two coupled $so(2n)$ cosets defined by the action 
\efourxi\ can be solved by a quantum group restriction of the 
${}_q d_{2n}^{(1)}$ affine Toda theory with 
$q= \exp (-i\pi/\beta_+^2 )$ or $q= \exp (-i\pi/\beta_-^2 )$.

\subsec{Four Coupled Minimal Models and ${}_q d_4^{(1)}$ }

The construction of the last section is special for $d_4^{(1)}$ since
here removing the node $\alv_{pert}$ leaves four decoupled $su(2)$ nodes. 
{}From our general approach, we expect that this case corresponds to
{\it four} coupled minimal models. 

Let $\alv_i, i=0,..,4$ denote the simple roots of $d_4^{(1)}$. The
central node is $\alv_{pert} = \alv_2$. We now chose
\eqn\efivei{
\gv = \gamma \sum_{i\neq 2} \alv_i , }
where $\gamma$ is the same as in \eIiv.  This leads to $c=4c_k$ 
where $c_k$ is the central charge \eIo\ of the $k-th$ minimal model
$\CC^{(k)}$. In order for each node $\alv_i , i \neq 2$ to represent 
a single copy of the minimal model $\CC^{(k)}$, one requires $\beta$ 
to be $\beta_+$ or $\beta_-$ as defined in \eIvii.  

The dimension of the perturbation is 
\eqn\efiveii{
\Delta_{pert} = \beta^2 + \sqrt{2} \beta \alv_2 \cdot \gv
= \beta^2 - 4\sqrt{2} \beta \gamma ~~~,  }
and one finds
\eqn\efiveiii{\eqalign{
\Delta_{pert} &= 4 \Delta_\sigma~~~, ~~~~~~~ {\rm for} ~~ \beta = \beta_+ 
~~~; \cr
\Delta_{pert} &= 4 \Delta_\sigt~~~, ~~~~~~~ {\rm for} ~~ \beta = \beta_-  
~~~.\cr
}}
Thus, the appropriate quantum group restriction  of the $d_4^{(1)}$ 
affine Toda theory with $q = \exp(-i\pi /\beta^2 )$ describes a model 
of four minimal conformal models all coupled at one point via the 
operators $\sigma,\sigt$. For $\beta = \beta_+$, the action is given by 
\eqn\efiveiv{
{\cal A} = \sum_{i=1}^4 {\cal A}_{\CC^{(k)}_i} + \lambda \int d^2 x 
~ \sigma_1 \sigma_2 \sigma_3 \sigma_4 , }
where the subscripts refer to which copy of $\CC^{(k)}$.  

An interesting case is $k=1$, which corresponds to four coupled Ising
models. They can be grouped into two pairs, each pair with $c=1$. We can
bosonize each pair with scalar fields $\phi_1$ and $\phi_2$. Then the 
action \efiveiv\ can be expressed in this case as 
\eqn\efivev{
{\cal A} = \inv{4\pi} \int d^2 x \( \sum_{i=1,2} \inv{2} 
\( \partial \phi_i\)^2  + \lambda \cos (\phi_1 /2 ) 
\cos (\phi_2 /2)\) ~~~. }
Since the interaction can be written as 
$\cos((\phi_1 + \phi_2)/2) + \cos((\phi_1 - \phi_2)/2 )$, 
one sees that this corresponds to two decoupled sine-Gordon models
each at $\beta^2/8\pi = 1/4$.\foot{Here $\beta$ is normalized in the
usual convention where $\beta^2/8\pi = 1/2$ is the free fermion point.}    
   
\newsec{Spectrum and S-matrices for Coupled Minimal Models} 

For the remainder of this paper we will be concerned only with the
models $\CM_k^{\sigma}$. In ref.\ref\rgand{G.M. Gandenberger, N.J. MacKay, 
and G.M.T. Watts, {\it Nucl. Phys.} {\bf B 465} (1996), 329} 
$\qa32$ invariant S-matrices were constructed\foot{We remark that one 
has the identifications $\a32 = a_3^{(2)}$ and $b_2^{(1)} \equiv
c_2^{(1)}$}. These are S-matrices in the unrestricted (vertex) form for
the fundamental multiplets of solitons.  There are two such fundamental
multiplets which transform in the $4$--dimensional vector $\{4\}$ 
and in the $6$--dimensional adjoint $\{6\}$ representations of $SO(4)_q$. 
The mass ratio of the two fundamental multiplets of solitons is\foot{The
parameters $\omega, \lambda$ introduced in \rgand\ take the values
$\omega= 1/(k+2)$, $\lambda = (k+6)/4(k+2)$.} 
\eqn\eIv{
\frac{M_{\{6\}}}{M_{\{4\}}} = 2 \cos \( \frac{\pi}{k+6} \) ~.}
In addition to these fundamental solitons there are scalar bound states 
and excited solitons depending on $k$ \rgand. As previously explained, 
the models $\CM_k^{\sigma}$ are described by quantum group (RSOS) 
restrictions of these S-matrices. The RSOS spectrum proposed in 
\rvays\ appears incomplete however. This will be evident below where 
we consider the Ising case at $k=1$.

\subsec{Ising Case and Relation to Sine-Gordon at $\beta^2/8\pi = 1/8$}

\def\I2{ {\rm Ising}^2_h }

The model $\CM_1^{\sigma}$ can be described by the action
\eqn\eIIi{
{\cal A}  = {\cal A}_{{\rm Ising}_1 } 
+ {\cal A}_{ {\rm Ising}_2 } 
+ \lambda \int d^2 x ~ \sigma_1 \sigma_2 ~~~,}
where $\sigma_{1,2}$ are the spin fields in copies ${\rm Ising}_{1,2}$. 
We will refer to the model \eIIi\ as ${\rm Ising}^2_h$. The presence of 
the coupling constant $\lambda$ destroys the critical fluctuations of 
the two individual models and the resulting system has a tendency to
acquire a net magnetization: its spectrum becomes then massive\foot{
In the following $\lambda$ is assumed to be positive. However, 
all the following conclusions hold independently from the sign of $\lambda$
since the sign of the coupling constant can be altered by changing the 
sign of one of the magnetization operators, say the one with index $1$: 
$\lambda \rightarrow -\lambda$ ; $\sigma_1 \rightarrow - \sigma_1$. 
${\cal A}_1$ is left invariant under this transformation since is the
action of the critical point.}. It is easy to predict the existence of 
kink excitations in the spectrum: in fact, there are two degenerate ground
states of the system \eIIi, one where both systems { have}
a positive total 
magnetization the other where the total magnetization is negative. The two 
ground states are related each other by the  
$Z_2$ symmetry\foot{The system presents another $Z_2$ symmetry related to
the exchange of the labels $1 \leftrightarrow 2$.} $\sigma_1 
\rightarrow - \sigma_1 ; \sigma_2 \rightarrow -\sigma_2$ and therefore 
there will be kink (antikink) excitations $K_{+-}$ ($K_{-+}$) interpolating
asymptotically between them (Figure 4). However, multi-kink configurations
can  only be  constructed in terms of a string of kink strictly followed by 
an antikink: $|...K_{+-} K_{-+} K_{+-} ...\rangle$. This means that the 
kinks of this system should behave actually like ordinary particles, as
will be indeed confirmed by the analysis which follows.    
\midinsert
\epsfxsize = 3in
\bigskip\bigskip\bigskip\bigskip
\vbox{\vskip -.1in\hbox{\centerline{\epsffile{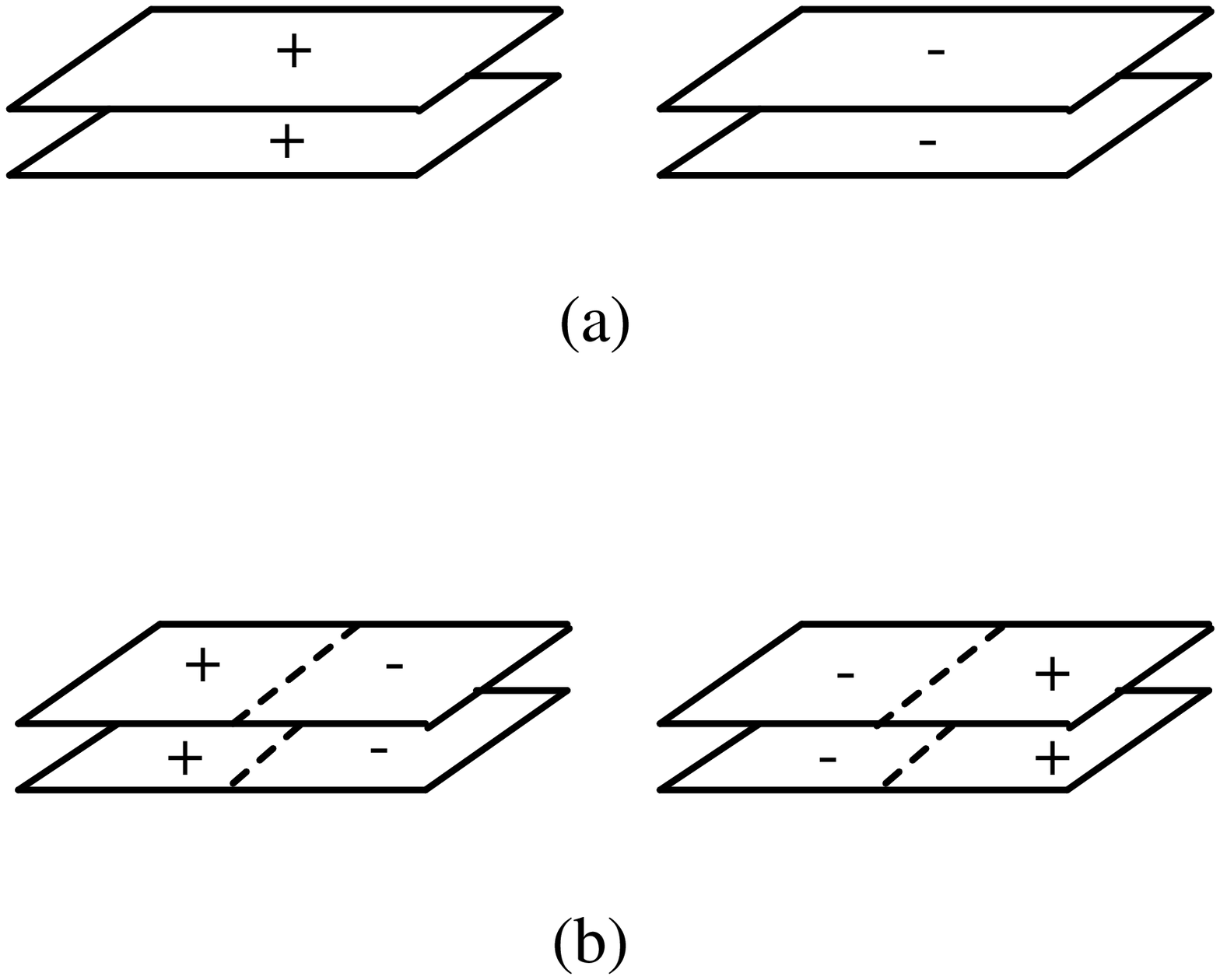}}}
\vskip .1in
{\leftskip .5in \rightskip .5in \noindent \ninerm \baselineskip=10pt
~~~~~Figure 2. Ground states (a) kink and antikink excitations.  
\smallskip}} \bigskip
\endinsert
\def\bh{\hat{\beta}}
\def\b8{{\beta^2/8\pi}} 

A simple argument relates the model \eIIi\ to the sine-Gordon theory
at the reflectionless point $\beta^2/8\pi = 1/8$. The sine-Gordon 
theory $SG_{\b8} $ is defined by the action
\eqn\sgaction{
{\cal A} = \inv{4\pi} \int d^2 x ~ 
\left[\inv{2} \(\partial \phi \)^2 +  \Lambda
\cos \bh \phi \right] , }
where $\bh = \beta/\sqrt{4\pi}$. {}From the Ising Majorana fermions
$\psi_{1,2}$, we can form a Dirac fermion $\psi_\pm = \psi_1 \pm i \psi_2$.
This is a $c=1$ CFT which can be bosonized by means of the formula 
$\psi_\pm = e^{\pm i \phi_L}$, where $\phi_L$ is the left-moving
component of $\phi$.   The operator $\sigma_1 \sigma_2$ has dimension
$1/4$, thus in the bosonized description it corresponds 
{  to $\cos(\phi/2)$, which} 
corresponds to $\beta^2/8\pi = 1/8$. Let us refer to the latter
theory as $SG_{1/8}$.

\def\stild{\tilde{\sigma} }

The above simple argument is not strictly correct since it ignores
the fact that the conformal field theory of ${\rm Ising}_1 \otimes 
{\rm Ising}_2 $, is not identical to that of a free Dirac fermion.  
Rather, it is an orbifold model at $R_{orb} = 1$, whereas the Dirac 
theory is a scalar field compactified on a circle with $R_{circle} =1$ 
\ref\gin{P. Ginsparg, {\it Nucl. Phys.} {\bf B 295} [FS21] (1988), 153; 
R. Dijkgraaf, E. Verlinde and H. Verlinde, {\it Comm. Math. Phys.} 
{\bf 115} (1988), 649.}. More generally, consider the theory $SG_\b8$.  
The potential in \sgaction\ has the symmetry $\phi \to \phi + 2\pi /\bh$, 
$\phi \to -\phi$. Thus the potential preserves the orbifold 
symmetries at $R_{orb} = 1/\bh$, and starting from $SG_\b8$ one can
easily define a perturbed orbifold version of it at this radius.
For $\b8 = 1/8$, $R_{orb} = 2$ and the resulting theory is the
$D_8^{(1)}$ theory \ref\rcor{H.W. Braden, E. Corrigan, P. E. Dorey
and R. Sasaki, {\it Phys. Lett.} {\bf B 227} (1989), 411.}\ref\klassen{T.
Klassen and E. Melzer, {\it Nucl. Phys.} {\bf B 338} (1990), 485.}.
The latter only differs from $SG_\b8$ by some signs in the S-matrices. 

\def\phit{\tilde{\phi}}

It turns out that the $SG_\b8$ theory is closely related to a second
perturbed orbifold CFT at the different radius $\tilde{R}_{orb} 
= R_{orb}/2$, and this is what corresponds to $\I2$.  To see this,
first redefine $\phi = \pi/2 - \tilde{\phi} $, so that the potential
in \sgaction\ becomes $V(\phit ) = \sin \bh \phit$.  The potential
now satisfies 
\eqn\epot{
V( \phit + 2\pi \tilde{R}_{orb} ) = - V(\phit) = V(-\phit) .}
Since in an orbifold, $\phit \sim - \phit$, we see that the potential
preserves the orbifold symmetry at $\tilde{R}_{orb} = R_{orb}/2 
= \sqrt{\pi/ \beta}$.  For $\b8 = 1/8$, this corresponds to 
$\tilde{R}_{orb} = 1$.    

We will resolve the distinction of $\I2$ from $SG_{1/8}$ and
$D_8^{(1)}$ by appealing to the formulation of the last section
based on $\qa32$. As we will show below, the spectrum is the same
as for $SG_{1/8}$, and the S-matrices again differ only by some signs.  
The final result can be anticipated more simply as follows. 
Let $s_1, s_2$ denote the excitations corresponding to the
$SG_{1/8}$ solitons, with mass $m_s$. Their S-matrices are  
\eqn\eIIiii{
S_{s_1 s_1} = S_{s_2 s_2} = \sigma' ~ S_{s_1 s_2 } 
= - \tilde{\sigma} ~ F_{1/7} (\theta)  F_{2/7} (\theta)  F_{3/7} (\theta)  , }
where
\eqn\eIIiv{
F_\alpha (\theta ) \equiv \frac{ \tanh \inv{2} ( \theta + i \pi \alpha ) } 
{\tanh \inv{2} ( \theta - i \pi \alpha )} 
~, } 
and $\theta$ is the rapidity variable, $E= m\cosh\theta$. 
The $SG_{1/8}$ model corresponds 
to $(\sigma' = 1 , \stild = 1 )$, whereas for the $D_8^{(1)}$ model 
$(\sigma' = -1, \stild = -1 )$. We claim that the $\I2$ model corresponds 
to the 3rd possibility
\eqn\eIIv{
\I2 : ~~~~~~~~ \sigma' = 1 , \stild = -1 . }
on the basis of the following argument. In the $SG_{1/8}$ model, 
there is a $U(1)$ symmetry under which $s_1 $ and $s_2$ are charge 
conjugate states. Crossing symmetry then implies $S_{s_1 s_1} 
=S_{s_1 s_2}$, i.e. $\sigma' = 1$. The breathers of the SG model  
are $s_1-s_2$ bound states, which implies a positive imaginary
residue in the corresponding poles of $S_{s_1 s_2} $, and this 
fixes $\stild = 1$. For the $D_8^{(1)}$ model on the other hand, 
since it is a perturbation of an orbifold theory, the $U(1)$ symmetry 
is broken to $Z_2$, and this allows $\sigma' = -1$ since $s_1 , s_2 $ are
no longer charge conjugated particles; the sign $\stild$ implies that 
the breathers continue to be $s_1 - s_2$ bound states. For the choice
\eIIv, the first breather is neither a $s_1 - s_1$, $s_2 -s_2$, nor 
$s_1 -s_2$ bound state, because none of these S-matrices have a positive
imaginary residue. We will show how this arises from the $\qa32$ 
description. 

The remaining S-matrices of the ${\rm Ising}^2_h$  model are the same as for
$SG_{1/8}$. There are 6 neutral excitations with mass 
\eqn\eIIvi{
m_a = m_1 \frac{ \sin \frac{a\pi}{14} }{ \sin \frac{\pi}{14} } 
~~~, 
~~~~~m_s = m_1 \inv{2 \sin \frac{\pi}{14} } , ~~~~~a = 1, 2,..,6}
and exact $S$--matrix amplitudes given by  
\eqn\eIIvii{\eqalign{
S_{ab} (\theta) &= \( \frac{|a-b|}{14} \) 
\[ \prod_{k=1}^{ {\rm min} (a,b) -1 }  
\( \frac{ |a-b| + 2k}{14} \) \]^2  \( \frac{a+b}{14} \)~;   
\cr
S_{as_1} (\theta) &= S_{a s_2} (\theta) = 
(-1)^a \prod_{k=0}^{a-1} \( \frac{7-a + 2k}{14} \) ~, 
\cr}}
where we have used $(\alpha) \equiv F_\alpha (\theta)$. 
Note that, as anticipated at
the beginning of this section, the kinks of this system behave indeed like 
ordinary particles, since their $S$--matrix can be entirely written in 
terms of the simple functions \eIIiv.

\def\1{[4]}

Now let us describe how the above result follows from the RSOS restriction
of $\qa32$. We denote the relevant $SO(4)_q$ representations
as $\{0\}, \{4\}, \{6\}$ for the singlet, vector, and adjoint
representations, respectively. The unrestricted S-matrix for the $\{4\}$  
can be written as \rgand\ref\rdel{G. W. Delius, M. D. Gould, and Y.-Z.
Zhang,  {\it Int. J. Mod. Phys.} {\bf A 11} (1996) 3415.}
\eqn\eIIviii{
S_{\{4\}\{4\}} (\theta) = \CF (\theta) ~ \tau_{21} \>  \hat{R}_{\{4\}\{4\}}
(x,q)  \> \tau_{12}^{-1} , }
where $\CF$ is a scalar factor, $\hat{R}_{\{4\}\{4\}} (x,q)$ is the
$R$-matrix for $\qa32$ multiplied by the permutation operator
$\bf P$, and $\tau_{12}$ is a gauge transformation. The $R$-matrix 
has the explicit form 
\eqn\eIIix{
\check{R}_{\{4\}\{4\}} = \check{P}_{\{9\}} 
+ \( \frac{1-xq^2}{x-q^2} \) \check{P}_{\{6\}} 
+ \( \frac{1+xq}{x+q} \) \check{P}_{\{0\}} ~~~,} 
where ${\bf P} \check{P}_{\rho}$ is a projector onto the 
$SO(4)_q$ representation $\rho$, and $x = \exp\( (k+6)\theta /(k+2) \) $.  
The scalar factor is given by
\eqn\efii{
\CF (\theta) = \frac{G_1 (\theta) G_{1-k/2} (\theta)}{G_0 (\theta) 
G_{-k/2} (\theta)} ~, }
where
\eqn\eG{
G_\alpha (\theta) = \prod_{j=1}^\infty 
\frac{ 
\Gamma \( \frac{k+6}{k+2} \( j - \frac{i\theta}{2\pi} \) 
- \frac{\alpha}{k+2} \) 
\Gamma \( \frac{k+6}{k+2} \( j - \frac{i\theta}{2\pi} \)
- \frac{4-\alpha}{k+2} \)} 
{\Gamma \( \frac{k+6}{k+2} \( j + \frac{i\theta}{2\pi} \)
- \frac{\alpha}{k+2} \) 
\Gamma \( \frac{k+6}{k+2} \( j + \frac{i\theta}{2\pi} \)
- \frac{4-\alpha}{k+2} \)} ~.} 
For the $\I2$ case, one must restrict the model at the root of unity
$q= - e^{i\pi/3}$, where $x = e^{7\theta/3}$. Specializing the formula\efii, 
 one finds
\eqn\eIIx{
\CF (\theta) = \( \frac{x+q}{xq+1} \) F_{1/7} (\theta )~. }

To perform the quantum group restriction we must examine the fusion
ring of $SO(4)_q$ at the above $q$. Recalling that $SO(4) = SU(2) \otimes
SU(2)$, let us label the $SO(4)_q$ representations as $(j, \tilde{j})$,
where $j$ denotes the spin $j$ representation of SU(2) with dimension 
$2j+1$, and similarly for $\tilde{j}$. The fundamental spinorial
representations of $SO(4)$ are the $(0,1/2), (1/2,0)$; there
are no fundamental multiplets of solitons in these representations. 
One also has $\{0\} = (0,0)$, $\{4\} = (1/2,1/2)$,  
$\{6\} = (0,1) \oplus (1,0)$, and $\{9\} = (1,1)$. At this root of unity, 
the $SU(2)$ fusion ring has a maximum spin $j=1/2$, and $(0) (1/2)  =
(1/2); (1/2) (1/2) = (0)$, where $(j) $ is an $su(2)$ spin $j$
representation. Since the $\{6\}$ of $SO(4)_q$ requires $j=1$, it is 
projected out of the spectrum. This leaves only the RSOS restriction of the
$\{4\}$, which is frozen and then behave like a scalar particle\foot{The
same freezing of degrees of freedom occurs when one restricts
the SG S-matrix to obtain the energy perturbation of the Ising model.}.  
The restriction leaves only the $\check{P}_{\{0\}}$ term in
$\check{R}_{\{4\}\{4\}}$.  Letting `$1$' denote the RSOS restriction of 
the fundamental soliton $\{4\}$, one then obtains the scalar S-matrix
\eqn\eIIxi{
S_{11} (\theta ) = F_{1/7} (\theta).} 
This is the S-matrix for the lightest sine-Gordon breather of $SG_{1/8}$. 
Let $m_1$ be the mass of this particle. Then closing the bootstrap for 
this particle leads to a total of six particles with masses and S-matrices 
given in \eIIvi\eIIvii. This is the spectrum proposed in \rvays.  

In the above analysis it is easy to overlook additional particles 
for the following reason. Though the $\{6\}$ is projected out, any
$\{6\}-\{6\}$ bound states which are scalars survive the restriction.  
These are the particles denoted as the $\qa32$ breathers $B_1^{(2)}, 
B_2^{(2)}$ in \rgand\foot{The breather $B_1^{(1)}$ is already
included as the second $SG_{1/8}$ breather.}. According to \rgand, the mass
of the particle $B_1^{(2)}$ is given by   
\eqn\eIIxii{
M_{B_1^{(2)}} = 2 m_1 \cos (\pi/7) \sin (3\pi/14) . } 
The mass of this particle can be identified with that of the
$SG_{1/8}$ soliton due the identity $4\cos(\pi/7) \sin (3\pi/14) = 
1/ \sin(\pi/14) $ (the latter identity is only valid due to the 14-th 
roots of unity involved). One can also check that the S-matrices involving
the particle $B_1^{(2)}$ computed in \rgand\ indeed correspond to those
in $SG_{1/8}$ with the assignment of signs \eIIv. Similarly,
the particle $B_2^{(2)}$ is identified with the 4-th $SG_{1/8}$ breather. 

So far we have the 6 breathers and one soliton of the $SG_{1/8}$ theory. 
A second soliton can be seen as necessary for the following reasons.  
The S-matrix for the scattering of the 1st and 6th $SG_{1/8}$ breathers 
is $S_{16} = F_{1/2} F_{5/14}$. The factor $F_{1/2}$ has a double pole
at $\theta = i\pi/2$. This corresponds to a ``bound state'' of mass
$M^2 = m_1^2 + m_6^2 = (2 m_s )^2$, i.e. to a state right at the threshold 
of a 2-soliton state. The fact that this pole indeed corresponds to a
2--soliton state is easily verified by checking that the S-matrices for this
``bound state'' with a particle $a$, as computed from the bootstrap, is
equal to $(S_{sa})^2$. Further reasons for this double degeneracy will be
given in the general case ahead.  

Due to the signs in \eIIv, the bound state structure of the model
$\I2$ is different from those  of $SG_{1/8}$ and $ D_8^{(1)}$.  
For the $\I2$ model, closing the bootstrap starting from the 
solitons $s_1$ and  $s_2$, and requiring a positive imaginary residue, 
leads to the 2nd, 4th, and 6th breathers. The odd breathers arise 
by closing the bootstrap starting from the 1st breather, which is 
viewed as a fundamental particle. {We remark that the sign
differences of the S-matrices  in \eIIiii \   do not change the TBA
analysis of the ultraviolet central charge, which  therefore reproduces
correctly
$c=1$ for all three cases, $SG_{1/8}$, $ D_8^{(1)}$
and $\I2$.}

\subsec{General Case of $\CM_k^\sigma$ } 

\def\jt{{\tilde{j}}}

The fundamental solitons in the $\{4\}$ and $\{6\}$ of $SO(4)_q$ become
RSOS kinks $K^{\{4\}}_{\rho_2 \rho_1}$ and $K^{\{6\}}_{\rho_2 \rho_1}$
with RSOS indices $\rho_2 , \rho_1$ labeling representations of 
$SO(4)_q$. The kinks $K^{\{6\}}$ are $K^{\{4\}}$ bound states
occurring at the bootstrap fusion pole $\theta = 2 i \pi /(k+6)$.
As in section 3, we use the decomposition $SO(4) = SU(2) \otimes SU(2)$ 
to label $SO(4)_q$ representations as $(j, \tilde{j})$, where $j, \tilde{j}
\in \CZ + 1/2$ are $SU(2)$ spins. The selection rule on the kink 
$K^{\rho_0}_{\rho_2 \rho_1}$ is that the representation $\rho_2$ must
appear in the tensor product $\rho_1 \times \rho_0$ within the fusion 
ring of $SO(4)_q$.  Hence, we will need the $SU(2)_q$ fusion ring at 
$q=q^{(h)} =-  \exp(i\pi/(k+2))$:
\eqn\eIVi{
( j_1 ) \times ( j_2 ) = 
\sum_{j = |j_1 - j_2|}^{{\rm min} 
(j_1 + j_2 , k-j_1 - j_2 ) } ( j ) , }
with 
$j \leq k/2$.  

Since $\{4\} = (1/2 ,1/2)$, the $\{4\}$ fundamental solitons become the
RSOS kinks:
\eqn\eIVii{
K^{\{4\}}_{(j_2 \jt_2)(j_1 \jt_1 )} (\theta ) , 
~~~~~~~ j_2 \in j_1 \times 1/2 , ~ \jt_2 \in \jt_1 \times 1/2 . }
Similarly, since $\{6\} = (0,1) \oplus (1,0)$ there are two kinds of
RSOS $\{6\}$ kinks:
\eqn\eIViii{\eqalign{
& K^{\{6\}}_{(j_2 \jt_2)(j_1 \jt_1 )} (\theta ) , 
~~~~~~~ j_2 \in j_1 \times 1 , ~ \jt_2 = \jt_1 
\cr
& \tilde{K}^{\{6\}}_{(j_2 \jt_2)(j_1 \jt_1 )} (\theta ) , 
~~~~~~~ j_2 = j_1  , ~ \jt_2 \in  \jt_1 \times 1 . 
\cr }}
The mass ratio of  $K^{\{6\}}, \tilde{K}^{\{6\}}$ to $K^{\{4\}}$ is given 
in \eIv.   

In addition to the above kinks there are breathers, which are scalar
kink-kink bound states.  Let $B_p^{(\{4\})}, ~ p=1,2,...$ denote the
$K^{\{4\}} - K^{\{4\}}$ bound state breathers, and $B_p^{\{6\}}$ the 
$K^{\{6\}} - K^{\{6\}}$ breathers. From the results in \rgand, one can
reach the following conclusions. As $k$ increases one enters a repulsive
regime wherein most breathers become unbound and disappear from the
spectrum. The $B_1^{\{4\}} , B_2^{\{6\}}$ breathers occur at the fusion 
pole $\theta = i\pi (2-k)/(k+6)$, whereas the $B_1^{\{6\}}$ breather occurs
at $4i\pi/(k+6)$. When $k=2$, for the mass of these breathers we have 
$M(B_1^{(\1)} ) = 2 M_{\{4\}}$ and $M(B_2^{\{6\}}) = 2 M_{\{6\}}$, 
thus $k=2$ is the threshold value for these breather state and they
disappear. The only remaining breather for all $k\geq 2$ is $B_1^{\{6\}}$, 
which we denote simply as $B$, with a mass given by 
\eqn\eIViv{
M_B = 4 M_{\{4\}} \cos \( \frac{\pi}{k+6} \) 
\cos \( \frac{2\pi}{k+6} \) ~. }
The S-matrix for this breather is 
\eqn\eIVv{
S_{BB} (\theta ) = F_{\frac{k+2}{k+6}} (\theta)  F_{\frac{k+4}{k+6}} (\theta) 
F_{\frac{k}{k+6}} (\theta) ~.} 
Since the breather $B$ can arise as a bound state of either 
$K^{\{6\}}$ or of $\tilde{K}^{\{6\}}$, we believe this breather
is doubly degenerate; certainly it is doubly degenerate in the 
Ising case where it is the $SG_{1/8}$ soliton. The threshold 
for the disappearance of the $B$ breather, i.e. when $M_B = 2 M_{\{6\}}$, 
occurs at $k=\infty$. 

Let us now come back to the problem of the $S$--matrix of the kink states. 
The S-matrices of the kink states are characterized by the exchange relation:
\def\sos{_{(j_3 \jt_3)(j_2 \jt_2)}^{(j_4\jt_4)(j_1\jt_1)}}
\eqn\eIVvi{
K^{\{4\}}_{(j_3\jt_3)(j_2 \jt_2)} (\theta_2 ) 
K^{\{4\}}_{(j_2\jt_2)(j_1 \jt_1)} (\theta_1 )
= \sum_{(j_4 \jt_4)} 
S\sos (\theta_2 - \theta_1) 
~ 
K^{\{4\}}_{(j_3\jt_3)(j_4 \jt_4)} (\theta_1 ) 
K^{\{4\}}_{(j_4\jt_4)(j_1 \jt_1)} (\theta_2 )
}
and similarly for the scattering involving $K^{\{6\}}$.  
The S-matrix in \eIVvi\ follows from \eIIviii\ with 
$x = \exp( 4 \lambda \theta)$, $q = -\exp(i\pi \omega)$
where we have defined $\lambda = (k+6)/4(k+2)$ and
$\omega = 1/(k+2)$:  
\eqn\eIVvii{
S\sos (\theta ) = \CF (\theta) 
\bigl[ P_{\{9\}} - 
\frac{ \sinh(2 \lambda \theta 
+ i\pi \omega )}{\sinh(2 \lambda \theta - i\pi \omega )} P_{\{6\}} 
+ 
\frac{ \cosh(2 \lambda \theta 
+ i\pi \omega/2  )}{\cosh(2 \lambda \theta - i\pi \omega/2 )} P_{\{0\}} 
\bigr]\sos} 
with $\CF$ the same as in \efii, and with the projectors in 
RSOS form. The latter form of the projectors can be expressed in 
terms of $q-6j$ symbols, as we now describe.  Clearly one has
\eqn\eIVviii{
P_{(j, \jt)} = P_j \tilde{P}_\jt ~, }
where $P_j$ is the projector onto the spin $j$ representation of
$SU(2)_q$ in the tensor product space $1/2 \times 1/2$, and similarly
for the second copy $\tilde{P}_\jt$. One also needs
\eqn\eIVix{
P_{\{0\}} = P_0 \tilde{P}_0 , ~~~~~
P_{\{6\}} = P_0 \tilde{P}_1 + P_1 \tilde{P}_0 , ~~~~~
P_{\{9\}} = P_1 \tilde{P}_1 ~. } 
The projectors $P_j$ in unrestricted vertex form have matrix elements
expressed in terms of q-Clebsch-Gordon coefficients:
\eqn\eIVx{
\langle 1/2,m_3; 1/2, m_4 | P_j | 1/2, m_1 ; 1/2, m_2 \rangle 
= \sum_m 
\langle 1/2,m_3; 1/2, m_4 | j,m\rangle_q \langle
j,m  | 1/2, m_1 ; 1/2, m_2 \rangle_q}
Going to the RSOS basis, and using the identity
\eqn\eident{\eqalign{
& \( \langle j,m|j_1,m_1;j_{23},m_{23} \rangle_q \)
\left\{ \matrix{j_1&j_2&j_{12} \cr j_3&j & j_23\cr} \right\}_q 
\cr
& ~ = \sum_{\matrix{m_2, m_3; \cr m_2 + m_3 = m-m_1}} 
\langle j_{23}, m_{23} | j_2 , m_2 ; j_3 , m_3 \rangle_q 
\langle j, m | j_{12} , m_{12} ; j_3 , m_3 \rangle_q 
\langle j_{12}, m_{12} | j_1 , m_1 ; j_2 , m_2 \rangle_q 
\cr } } 
one obtains the simple result \ref\rpasq{V. Pasquier, {\it Commun. Math.
Phys.} {\bf 118} (1988), 355.}\ref\rkolya{A. N. Kirillov and N. Yu.
Reshetikhin, Advanced Series in Mathematical Physics vol. 7 (1989), 
V. G. Kac, ed., World Scientific.} 
\eqn\eIVxi{
\( P_j \)_{j_3 j_2}^{j_4 j_1} = 
\left\{ \matrix{ 1/2 & 1/2 & j\cr j_3 & j_1 & j_4 \cr } \right\}_q 
\left\{ \matrix{ j_3 & 1/2 & j_2\cr 1/2  & j_1 & j \cr } \right\}_q 
.} 
The $q-6j$ symbols can be found in \rkolya\rbl. The complete S-matrix 
follows from \eIVvii\eIVviii\ and \eIVix, along with the evident 
relation
\eqn\eevid{
\( P_{(j, \jt)} \)\sos =
\( P_j \)_{j_3 j_2}^{j_4 j_1} 
\( \tilde{P}_\jt \)_{\jt_3 \jt_2}^{\jt_4 \jt_1} . } 
The analog of the formula \eIIix\ involving $\{6\}$ fundamental 
solitons is unknown. However, the kinks $K^{\{6\}}, \tilde{K}^{\{6\}}$ 
are bound states of the kinks $K^{\{4\}}$ occurring at the fusion pole
$\theta =2 i \pi/(k+6)$. Therefore, the S-matrices involving
the $\{6\}$-kinks can in principle be computed from bootstrap 
fusion\foot{The spectrum proposed in \rvays\ does not contain 
the kinks $K^{\{6\}}, \tilde{K}^{\{6\}}$ nor the breather $B$. 
Also, the conjectured S-matrices for the $K^{\{4\}}$ kinks were 
constructed by borrowing RSOS solutions of the Yang-Baxter equation 
which define certain lattice statistical mechanics models in 
\ref\rjim{M. Jimbo, T. Miwa and M. Okada, {\it Commun. Math. Phys.} 
{\bf 116} (1988), 507.}. We have not checked if they agree with 
the S-matrices constructed here.}.

\subsec{The $k=\infty$ limit} 

When $k=\infty$, the model $\CM_k^{\sigma}$ corresponds to two level-1
$SU(2)$ current algebras coupled via their primary field in the spin 
$1/2$ representation of dimension $1/4$. Denoting the latter by 
${\bf \Phi}^{1/2}$, the action \eIii\ becomes 
\eqn\eIVA{
{\cal A} = {\cal A}_{su(2)_1} + {\cal A}_{su(2)_2} 
+ \lambda \int d^2 x ~ {\bf \Phi}^{1/2}_1   {\bf \Phi}^{1/2}_2 , } 
where $su(2)_{1,2}$ refers to the copies $1$ and $2$ of the current 
algebra. Above, 
\eqn\eIVB{
{\bf \Phi}^{1/2}_{1,2} = \sum_{m = \pm 1/2} 
{\bf \phi}^{(1/2,m)}_{1,2} \bar{\bf \phi}^{(1/2,-m)}_{1,2} , }
where ${\bf \phi}^{(1/2,m)}$, and $\bar{{\bf \phi}}^{(1/2,m)}$ 
are the left and right moving factors. 

The current algebras can each be bosonized with a { scalar field 
$\varphi_{1,2}$.}  The primary fields have the representation: 
\eqn\eIVBb{
{\bf \phi}_{1,2}^{(1/2,m)} = e^{i \sqrt{2} m \varphi_{1,2} }, 
~~~~~~~  
\bar{{\bf \phi}}_{1,2}^{(1/2,m)} = e^{- i \sqrt{2} m \bar{\varphi}_{1,2} }
~. }
Thus,
\eqn\eIVC{
{\bf \Phi}^{1/2}_1   {\bf \Phi}^{1/2}_2 = 
\( e^{i\eta_1 /\sqrt{2} } + e^{-i\eta_1 /\sqrt{2}} \) 
\( e^{i\eta_2 /\sqrt{2} } + e^{-i\eta_2 /\sqrt{2}} \) 
~, }
where
$\eta_{1,2} = \varphi_{1,2} + \bar{\varphi}_{1,2}$ are local scalar fields. 
Finally, the interaction can be expressed in terms of fermion bilinears:
\eqn\eIVD{
{\bf \Phi}^{1/2}_1   {\bf \Phi}^{1/2}_2 =
\psi_+ \bar{\psi}_- + \bar{\psi}_+ \psi_- + 
\psi'_+ \bar{\psi'}_- + \bar{\psi}'_+ \psi'_- , }
with 
\eqn\eIVE{
\psi_+ \bar{\psi}_- = e^{i (\eta_1 + \eta_2)/\sqrt{2} }~~~, 
~~~~~
\psi'_+ \bar{\psi}'_- = e^{i (\eta_1 - \eta_2)/\sqrt{2} }~.}
Since the fermions are complex, combined together they correspond to 
$4$ real fermions. The interaction simply gives each real fermion
the same mass. Thus, as $k\to \infty$, the model $\CM_k^{\sigma}$ 
becomes the free field theory of $4$ real massive fermions with
$SO(4)$ symmetry. This result is closely related to the lattice
model results obtained in\rtvel\foot{For
 the  
lattice model considered in \rtvel\  the $SO(4)$ symmetry is broken 
to $\CZ_2 \times SU(2)$ which leads to a triplet and a singlet of
fermions of different mass.}.  

The above result arises from the restricted $\qa32$ symmetry in the 
following way. Firstly, as $k\to \infty$, in the $(j, \jt)$ labeling of
$SO(4)$ representations, we have $j_{\rm max} = \infty$; this implies that
the RSOS S-matrices are unrestricted (SOS) and by a change of basis can
be brought back to unrestricted vertex form. Secondly, since
$M_{\{6\}} = 2 M_{\{4\}}$, $k=\infty$ is the threshold for the disappearance
of the fundamental solitons of mass $M_{\{6\}}$. Also, $M_B = 2 M_{\{6\}}$, 
so that the breather $B$ also disappears. This leaves a 4-plet of 
solitons transforming under the undeformed vector of $SO(4)$, and these
are the 4 real fermions. Finally, since $q\to -1$, the $\qa32$ is
undeformed. It is known that $2n$ free massive fermions has an undeformed
$a_{2n-1}^{(2)}$ symmetry algebra\ref\rsot{E. Abdalla, M.C.B. Abdalla, 
G. Sotkov and M. Stanishkov, {\it Int. J. Mod. Phys.} {\bf A 10} (1995), 
1717.}\ref\rlec{A. LeClair, {\it Nucl. Phys.} {\bf B 415} (1994), 734.}, 
thus the S-matrices above must become free as $k\to \infty$.   
 
\newsec{Conclusions}

In this paper we have studied the on--shell dynamics of coupled 
conformal field theories under the constraint of the integrability 
for the inter-layer coupling. A general framework for this kind of 
models is provided by the reduction of Affine Toda Field Theory 
{associated  with } particular Dynkin diagrams. These are the 
Dynkin diagrams of the affine algebra $\hat{g}$ which have the property
that upon removing one of its nodes, one is left with two decoupled 
Dynkin diagrams $g_1$ and $g_2$ of finite dimensional simply laced  
Lie algebras: the latter represent then { the Lie algebras from
which tensor product of  two minimal models is 
constructed using the  coset
construction. The removed node, on the other hand,
specifies a particular integrable coupling
between these two  minimal models.} An interesting model of this 
class is represented by the two--layer Ising model coupled by the 
magnetic operators: this model has been analyzed both in terms of 
the RSOS restriction of $\qa32$ as well as in terms of a bosonization 
scheme related to the Sine-Gordon model. It would be interesting to 
pursue further the analysis of this model as well as of the others by
computing their form factors and their correlation functions: 
quantities particularly interesting in this respect would be the
correlators involving operators living on the two different planes, 
as for instance the correlator $\langle\sigma_1(x) \sigma_2(y) \rangle$ 
for the two--layer Ising model. 
Also it is clear that there are other interesting examples of the
Dynkin gymnastics used in this paper. 
Finally, at a speculative level,  one might ask if
our integrable analysis  of  two (or four) coupled minimal models, 
discussed in this paper,
could be some first step in  understanding 
$N$ coupled conformal field theories.   If this could be
done systematically, one  may perhaps hope to be able to learn
something about three dimensional theories.

\bigskip

\centerline{\bf Acknowledgments}

We would like to thank G. Delfino, G. Delius,  G. Gandenberger,  
 F. Lesage, H. Saleur,  G. Sierra, and P. Simonetti for discussions. 
{A.Lec. and G.M.} would also  like 
to thank the Institute for Theoretical Physics 
in Santa Barbara and the organizers of the program {\it Quantum Field Theory 
in Low Dimensions: From Condensed Matter to Particle Physics} which was
held there, for the warm hospitality.   
This research is supported in part by the National Science Foundation
under Grant. No. PHY94-07194.  {{A.Lec.} is 
supported in part by the National
Young Investigator Program of the NSF,  {A.W.W.L} by the A.P. Sloan
Foundation, and G.M. by  
EC TMR Programme, grant FMRX-CT96-0012.}

\listrefs
\bye